\documentclass[twocolumn,epjc3]{svjour3}

\usepackage[T1]{fontenc}
\usepackage[utf8]{inputenc}
\usepackage{color}
\usepackage{graphicx}
\usepackage{amsmath}
\usepackage{amsfonts}
\usepackage{mathdots}
\usepackage{mathrsfs} 
\usepackage{dsfont} 
\usepackage{stmaryrd}
\usepackage{enumerate}
\usepackage{hyperref}
\hypersetup{colorlinks=true,
            linkcolor=blue,
            citecolor=blue,
            filecolor=green,
            urlcolor=cyan 
}

\usepackage[numbers,sort&compress]{natbib}
\usepackage{scalefnt}
\usepackage{txfonts}
\usepackage{microtype}

\journalname{Eur. Phys. J. A}
\renewcommand{\email}[1]{e-mail: \href{mailto:#1}{#1}}

\newcommand{\bra}[1]{\langle #1 \vert}
\newcommand{\ket}[1]{\vert #1 \rangle}
\newcommand{\scal}[2]{\langle #1 \vert #2 \rangle}
\newcommand{\elma}[3]{\bra{#1} #2 \ket{#3}}

\newcommand{\TAURUS}{\textsf{TAURUS}}
\newcommand{\TAURUSvap}{$\textsf{TAURUS}_{\textsf{vap}}$}
\newcommand{\TAURUSpav}{$\textsf{TAURUS}_{\textsf{pav}}$}
\newcommand{\TAURUSmix}{$\textsf{TAURUS}_{\textsf{mix}}$}

\graphicspath{{.}{figures/}}

\begin{document}

\title{Symmetry-projected variational calculations with the numerical suite \TAURUS}
\subtitle{II. Configuration mixing of symmetry-projected reference states}

\author{B.~Bally\thanksref{ad:esnt,em:bb} \and  T.~R.~Rodríguez\thanksref{ad:dft,ad:ucm,ad:ciaff,em:trr}}
\date{Received: \today{} / Accepted: date}

\thankstext{em:bb}{\email{benjamin.bally@cea.fr}}
\thankstext{em:trr}{\email{tomasrro@ucm.es}}

\institute{
\label{ad:esnt}
ESNT, IRFU, CEA, Universit\'e Paris-Saclay, 91191 Gif-sur-Yvette, France 
\and
\label{ad:dft}
Departamento de F\'isica Te\'orica, Universidad Aut\'onoma de Madrid, E-28049 Madrid, Spain  
\and 
\label{ad:ucm}
Departamento de Estructura de la Materia, F\'isica T\'ermica y Electr\'onica, Universidad Complutense de Madrid, E-28040 Madrid, Spain
\and
\label{ad:ciaff}
Centro de Investigaci\'on Avanzada en F\'isica Fundamental-CIAFF-UAM, E-28049 Madrid, Spain 
}

\maketitle
%
%
\begin{abstract}
We present the numerical codes \TAURUSpav~and \TAURUSmix~that, combined, perform the configuration mixing of symmetry-projected real general Bogoliubov quasiparticle states represented in a spherical harmonic oscillator basis. The model space considered is invariant under spatial and isospin rotations but no specific set of orbits is assumed such that the codes can carry out both valence-space and no-core calculations. 
In addition, no number parity is assumed for the Bogoliubov quasiparticle states such that the codes can be used to describe even-even, odd-even and odd-odd nuclei. 
To demonstrate the potential of the codes, we perform an example no-core calculation of $^{24}$Mg using a modern microscopic interaction.
\end{abstract}
%
%
\section*{PROGRAMS SUMMARY}
\begin{description}[font=\normalfont\itshape]
  \item[Program title:] 1) \TAURUSpav
  \item[\phantom{Program title:}] 2) \TAURUSmix
  \item[License:] GNU General Public License version 3 or later
  \item[Programming language:] Fortran 2008
  \item[DOI:] 1) \href{https://doi.org/10.5281/zenodo.10420261}{10.5281/zenodo.10420261}
  \item[\phantom{DOI:}] 2) \href{https://doi.org/10.5281/zenodo.10419912}{10.5281/zenodo.10419912}
  \item[Repository:] 1) \href{https://github.com/project-taurus/taurus\_pav}{github.com/project-taurus/taurus\_pav} 
  \item[\phantom{Repository:}] 2) \href{https://github.com/project-taurus/taurus\_pav}{github.com/project-taurus/taurus\_mix}
  \item[Nature of problem:]
   The configuration mixing of symmetry-projected Bogoliubov quasiparticle states, in the spirit of the Projected Generator Coordinate Method, is a variational beyond-mean-field method that is well suited to tackle strong collective correlations in atomic nuclei.
   Nevertheless, no sophisticated and publicly available numerical code exists that can 
   handle general Hamiltonians as constructed, for example, in state-of-the-art effective field theories.
  \item[Solution method:]
   We propose two numerical codes that, combined, can be used to perform the configuration mixing of parity, particle-number and total-angular-momentum projected real general
   Bogoliubov quasiparticle states represented in a spherical harmonic oscillator basis. 
   The codes offer several degrees of parallelization and make use of fast mathematical routines from the \textsf{BLAS} and \textsf{LAPACK} libraries.
  \item[Additional comments:]
  The codes \TAURUSpav\ and \TAURUSmix\ are part of the numerical suite \TAURUS\ that permits advanced symmetry-projected variational calculations.
\end{description}
%
%
\section{Introduction}
\label{sec:intro}

The mean-field approximation is a simple, yet powerful, concept that has been used for decades to investigate the structure of atomic nuclei \cite{Bender03a,RS80a}. In particular, this approximation reaches its full potency when one allows the mean field to break one or several symmetries of the nuclear Hamiltonian, as such a  breaking permits to incorporate important effects of many-body correlations within a simple one-body picture. A notable example of symmetry-breaking mean-field theory is the unrestricted Hartree-Fock-Bogoliubov (HFB) method \cite{Bender03a,RS80a} within which one searches for the trial state yielding the lowest expectation value of the Hamiltonian exploring  the variational space of Bogoliubov quasiparticle product states. For instance, the energy minimum obtained in realistic calculations of open-shell nuclei often corresponds to a spatially deformed one-body density that breaks the rotational invariance \cite{Bender06a,Rodriguez15a,Scamps21a}, which can be understood as a manifestation of a Jahn-Teller-like effect \cite{Reinhard84a}.

In spite of its advantages, the mean-field approximation is too elementary to capture all the intricate correlations emerging in the complex system of interacting nucleons that is the atomic nucleus. In addition, the lack of good quantum numbers for the approximate states, due to the breaking of nuclear Hamiltonian symmetries, often makes dubious the direct comparison of computed quantities with experimental data. The Projected Generator Coordinate Method (PGCM) \cite{Hill53a,Griffin57a,Bender03a,RS80a} is a variational beyond-mean-field technique that addresses both problems simultaneously and consistently. In this approach, the trial wave functions are constructed as linear superpositions of several symmetry-projected references states that differ by a set of collective coordinates. In this work, the reference states are taken to be Bogoliubov quasiparticle states obtained through a series of constrained HFB minimizations \cite{Bally21b}. The coefficients in the superposition are variational parameters that are determined by minimizing the expectation value of the Hamiltonian, which in a discrete basis is equivalent to solving a generalized eigenvalue problem. 
In the end, the approximate states built in such a fashion have good quantum numbers and incorporate further collective correlations. Ultimately, the PGCM can be interpreted as a general restoration of symmetries within which both the modulus and the phase of the order parameter associated with the breaking of the symmetry are explored \cite{Duguet10a}.

Along the years, the PGCM has been employed extensively in nuclear structure investigations using various representations of the nuclear Hamiltonian: phenomenological energy density functionals \cite{Bonche90a,Bender06a,Bally14a,Bally22b,Bally23a,Rodriguez15a,Egido16a,Robledo18a,Yao10a,Marevic18a,Marevic19a,Zhou23a}, empirical valence-space interactions \cite{Wang18a,Bally19a,Sanchez21a,Shimizu21a,Dao22a} and realistic interactions obtained from chiral Effective Field Theory ($\chi$EFT) \cite{Yao20a,Yao21a,Frosini22a,Frosini22b,Frosini22c,PorroPHD}. The computer codes used to perform these advanced calculations, however, rarely have been publicly shared. A notable exception is the software \textsf{HFODD} \cite{Dobaczewski21a}.

To remedy the situation, in this article, we present the numerical codes \TAURUSpav\ and \TAURUSmix~that can be employed to perform PGCM calculations based on real general Bogoliubov quasiparticle reference states projected simultaneously onto good parity, particle numbers and total angular momentum. The Bogoliubov states are represented in a spherical harmonic oscillator (SHO) basis invariant under spatial and isospin rotations but no specific set of orbits is assumed such that the codes can carry out both valence-space and no-core calculations. 
In addition, no number parity is assumed for the Bogoliubov states such that the code can be used to describe even-even, odd-even and odd-odd nuclei. The numerical codes \TAURUSpav\ and \TAURUSmix\ are part of the numerical suite \TAURUS,  the first component of which was previously published \cite{Bally21b}, and were already used to perform several realistic calculations in recent years \cite{Bally19a,Duguet20a,Sanchez21a,Romero21a,Yao20a,Frosini22b,Frosini22c}.
 
The article is organized as follows. In Sect.~\ref{sec:theory}, we give an account of the theoretical framework. Then, in Sect.~\ref{sec:example}, we perform an example calculations using a $\chi$EFT-based Hamiltonian. 
In Sect.~\ref{sec:conclu}, we present a summary of this work as well as perspectives for the future developments of the numerical suite \TAURUS. Finally, Sects~\ref{sec:structpav} and \ref{sec:structmix} focus on the most
relevant technical details of the programs \TAURUSpav~and \TAURUSmix.
%
%
\section{Theoretical framework}
\label{sec:theory}

\subsection{Preliminary definitions}
\label{sec:theo:prel}

First, let us consider a symmetry group $\mathcal{G}$ with elements $\Omega$ and its unitary representation $R(\Omega)$ in the many-body Hilbert space. Generically, we call $\Omega$ an \emph{angle} and  $R(\Omega)$ a \emph{rotation}.
In this work, we consider $\mathcal{G}$ to be either a finite or a compact Lie group and note $v_\mathcal{G}$ either the order or the volume of the group. In this case, the irreducible representations (irreps) of $\mathcal{G}$ are all finite dimensional \cite{Hamermesh62a} and we note $d_\lambda$ the dimension of irrep $\lambda$. Finally, the matrix representation of the rotation $R(\Omega)$ for irrep $\lambda$ is labelled $D^\lambda_{ij}(\Omega)$ with $i,j \in \llbracket 1, d_\lambda \rrbracket^2$. 

Then, let us consider a (set of) parameter(s) $q$ that measures the magnitude of symmetry breaking of a wave function and that is such that $q=0$ for a symmetry-conserving state and $|q| > 0$ for a symmetry-broken state. Generically, we call $q$ a \emph{deformation}. Usually, $q$ is taken to be the expectation value(s) of a (set of) general one-body operator(s) related to the symmetry at play. 

For example, considering the group $SU(2)$ associated with the rotational invariance in the full space, i.e., including spatial and spin degrees of freedom, the rotations can be be parametrized by the so-called Euler angles $\alpha_J,\beta_J,\gamma_J \in \left[0,2 \pi \right] \times \left[0, \pi \right] \times \left[0, 4 \pi \right]$ \cite{Varshalovich88a} and the amount of symmetry breaking can be measured by non-vanishing expectation values of the multipole operators $Q_{lm} \equiv r^l Y_{lm}$, with $r$ being the position and $Y_{lm}$ a spherical harmonic of degree $l$ and order $m$.   

In general, the total symmetry group considered can be decomposed as a direct product of subgroups and the angles and parameters can be written as sets of coordinates, i.e., typically we have $\Omega \equiv (\Omega_1, \ldots, \Omega_m)$ and $q \equiv (q_1, \ldots, q_n)$.

\subsection{Variational ansatz}
\label{sec:theo:ans}

Let us consider a set of reference states $ \Sigma \equiv \left\{ \ket{\Phi_a}, a \right\}$. For the sake of generality no assumption is made at this point regarding the precise nature and properties of the reference states. 

To build approximations to the eigenstates of the nuclear Hamiltonian, we consider the variational ansatz
\begin{equation}
 \label{eq:ansatz}
  \ket{\Psi^{\lambda i}_{\Sigma \epsilon}} = \frac{d_\lambda}{v_G} \sum^{\Sigma}_{a} \sum^{d_\lambda}_{j=1} f^{\lambda j}_{{\Sigma a \epsilon}} \int d\Omega D^{\lambda *}_{ij}(\Omega) R(\Omega) \ket{\Phi_a} , 
\end{equation}
where $\left\{ f^{\lambda j}_{{\Sigma a \epsilon}} \right\}$ are parameters that are determined through the minimization of the total energy, 
\begin{equation}
 \label{eq:var}
  \delta \left( \frac{\elma{\Psi^{\lambda i}_{\Sigma \epsilon}}{H}{\Psi^{\lambda i}_{\Sigma \epsilon}}}{\scal{\Psi^{\lambda i}_{\Sigma \epsilon}}{\Psi^{\lambda i}_{\Sigma \epsilon}}} \right) = 0 ,
\end{equation}
performed independently for each irrep $\lambda$. 
The index $\epsilon$ is then used to label the different energies in the spectrum thus obtained. 
When working with a finite set $\Sigma \equiv \left\{ \ket{\Phi_a} , a \in \llbracket 1, n_\Sigma \rrbracket \right\}$, which in practice is always the case, the variational equation \eqref{eq:var}, commonly named Hill-Wheeler-Griffin equation, can be recasted as a generalized eigenvalue problem (GEP).

The weighted integral of rotations over all angles, which can be interpreted as a projection operator \cite{Bally21a,Sheikh21a,Tanabe05a}
\begin{equation}
\label{eq:proj}
  P^{\lambda}_{ij} \equiv \frac{d_\lambda}{v_G}  \int d\Omega D^{\lambda *}_{ij}(\Omega) R(\Omega) , 
\end{equation}
assures that the ansatz of Eq.~\eqref{eq:ansatz} has the good quantum numbers associated with the symmetry group $\mathcal{G}$.

The method introduced here is very general and can be interpreted as a generic configuration mixing of arbitrary, possibly non-orthogonal, reference states. 
In particular, this approach allows us, starting from a set of arbitrary reference states, $\Sigma$, to naturally build a spectrum of more correlated states that respect the symmetries of the nuclear Hamiltonian. 
Having access to the wave functions, and given a general operator $O$ of interest, it is then also possible to compute the transition matrix elements $\elma{\Psi^{\lambda' i'}_{{\Sigma \epsilon'}}}{O}{\Psi^{\lambda i}_{\Sigma \epsilon}}$, which by construction obey the appropriate selection rules.

\subsection{Projected Generator Coordinate Method}
\label{sec:theo:coll}

A specific realization of the approach described above is the Projected Generator Coordinate Method \cite{Bender03a,RS80a}. 
The original idea suggested by Hill, Wheeler and Griffin \cite{Hill53a,Griffin57a}, was to build a linear superposition using a family of states that depend on a so-called (collective) generator coordinate. But this concept can be naturally extended to include also the symmetry restoration first proposed by Peierls and Yoccoz \cite{Peierls57a}.

\begin{figure}[t!]
\centering  
  \includegraphics[width=1.00\columnwidth]{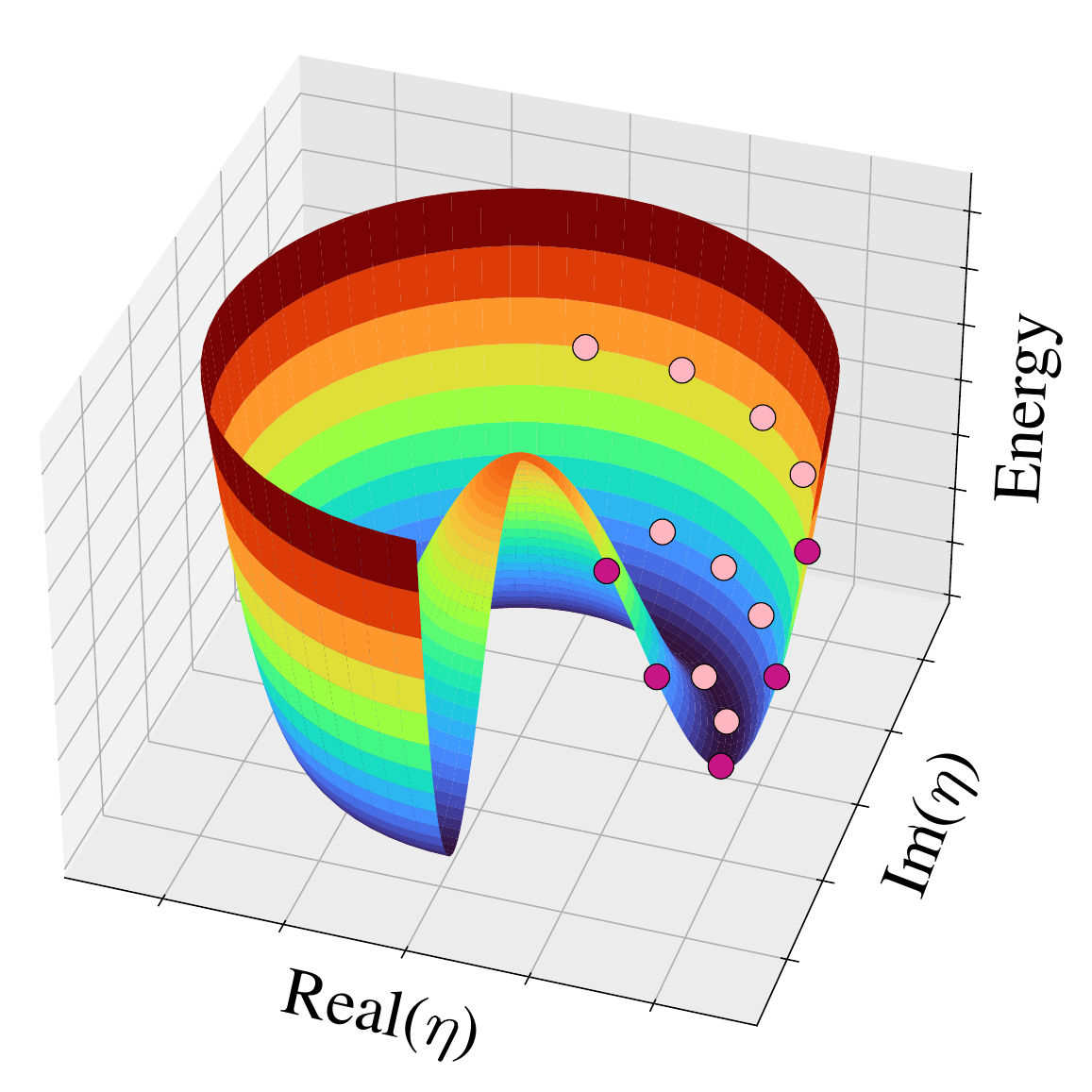} 
\caption{
\label{fig:manifold}
(Color online)
Schematic representation of the energy surface as a function of the order parameter $\eta = q e^{i\Omega}$. The dark pink circles represent the discretization of the surface along the modulus of $\eta$. The light pink circles represent the discretization of the surface along the argument of $\eta$. 
Within the PGCM, the variational ansatz considered is a linear superposition of reference states generated by exploring simultaneously those two dimensions
}
\end{figure}

Within that formulation of the PGCM, we consider as generator coordinate the order parameter of the group $\mathcal{G}$, namely $\eta \equiv \left( q , \Omega \right)$. In Eq.~\eqref{eq:ansatz}, this corresponds to consider the set $\Sigma \equiv \left\{ \ket{\Phi_a (q)}, a , q \right\}$,\footnote{While the parameter $q$ is \emph{a priori} continuous, we remark that is always possible to approximate this family of states by a countable set \cite{Broeckhove79a}.} where, for the sake of generality, we kept the index $a$, and we have $\elma{\Phi_a(q)}{Q}{\Phi_a(q)} = q$ for a certain (unspecified) operator $Q$. The method then allows us to fully explore the manifold of order parameters as schematically represented in Fig.~\ref{fig:manifold}.

The PGCM is very efficient at capturing the physics of strong (or static) correlations such as collective deformations or pairing. Conversely, the method is not well suited to include weak (or dynamical) correlations, which typically involve a large number of a few particle-hole excitations on top a given reference state. While the ansatz of Eq.~\eqref{eq:ansatz} is in theory general enough to include such configurations, this would lead in practice to consider a set $\Sigma$ of large dimension and, therefore, to a prohibitive computational cost. A manner to circumvent this deficiency was recently proposed through the formulation of the PGCM + Perturbation Theory \cite{Frosini22a,Frosini22b,Frosini22c,Duguet23a}.

\subsection{Implementation}
\label{sec:theo:imple}

A precise account of our theoretical framework was already given in our previous article dedicated to the numerical code \TAURUSvap~\cite{Bally21b} and, therefore, will not be repeated here. Instead, we will focus mainly on the new elements specific to the codes \TAURUSpav~and \TAURUSmix.

\subsubsection{Reference states}

The reference states $\ket{\Phi_a}$ are taken to be real general Bogoliubov quasiparticle states expanded on a SHO single-particle basis. Such states are characterized by their (real) matrices $(U_a,V_a)$ that define the so-called Bogoliubov transformation between the set of quasiparticle creation/annihilation operators $\left\{ \beta_{ap} ; \beta^{\dagger}_{ap} \right\}$, for which $\ket{\Phi_a}$ is a vacuum, and the set of single-particle creation/annihilation operators in the SHO basis $\left\{ c_{p} ; c^{\dagger}_{p} \right\}$, i.e., 
\begin{equation}
\begin{pmatrix}
\beta_a \\
\beta_a^{\dagger}
\end{pmatrix}
=
\begin{pmatrix}
U_a^\dagger & V_a^\dagger \\
V_a^T & U_a^T
\end{pmatrix} 
\begin{pmatrix}
c \\
c^{\dagger}
\end{pmatrix} .
\end{equation}
The SHO model space considered is invariant under spatial and isospin rotations but no specific set of orbits is assumed such that the codes can carry out both valence-space and no-core calculations. Also, we emphasize that no number parity is assumed for the Bogoliubov quasiparticle states such that the codes can be used to describe even-even, odd-even and odd-odd nuclei.

\subsubsection{Projection operators}

The reference states can be projected onto any subset of the following quantum numbers:  the number of protons $Z_0$, the number of neutrons $N_0$, the number of nucleons $A_0$, the total angular momentum $J_0$ and its third component $M_{J0}$, and the parity $\pi_0$. This implies that, in the most general case, we consider the irrep $(\lambda;j) \equiv (Z_0,N_0,A_0,J_0,\pi_0;M_{J0})$. In that case, we can decompose the global symmetry projection operator $P^{\lambda}_{ij}$ of Eq.~\eqref{eq:proj} in terms of separate projection operators 
\begin{subequations}
\begin{align}
  P^{\lambda}_{ij} &\equiv P^{Z_0} P^{N_0} P^{A_0} P^{J_0}_{M_{J0} K_{J0}} P^{\pi_0}, \\
  P^{Z_0} &= \frac{1}{2\pi} \int_{0}^{2\pi} d\varphi_Z \, D^{Z_0 *} (\varphi_Z) \, R^{Z} (\varphi_Z) , \\
  P^{N_0} &= \frac{1}{2\pi} \int_{0}^{2\pi} d\varphi_N \, D^{N_0 *} (\varphi_N) \, R^{N} (\varphi_N) , \\
  P^{A_0} &= \frac{1}{2\pi} \int_{0}^{2\pi} d\varphi_A \, D^{A_0 *} (\varphi_A) \, R^{A} (\varphi_A), \\
  P^{J_0}_{M_{J0} K_{J0}} &= \frac{2J_0 + 1}{16\pi^2} \int_{0}^{2\pi} d\alpha_J \int_{0}^{\pi} 
          d\beta_J \sin({\beta_J}) \int_{0}^{4\pi} d\gamma_J \\ 
          &\phantom{=}\, D^{J_0 *}_{M_{J0} K_{J0}} (\alpha_J,\beta_J,\gamma_J)   \, R^J(\alpha_J,\beta_J,\gamma_J) , \nonumber \\
  P^{\pi_0} &= \frac{1}{2} \sum_{\varphi_\pi = 0, \pi} D^{\pi_0 *} (\varphi_\pi) \, R^{\pi} (\varphi_\pi) .
\end{align}
\end{subequations}
The matrix elements of the irreps have the expressions 
\begin{subequations}
\begin{align}
  D^{\lambda *}_{ij}(\Omega) &\equiv  D^{Z_0 *} (\varphi_Z)  D^{N_0 *} (\varphi_N)  D^{A_0 *} (\varphi_A) \\
     &\phantom{==} D^{J_0 *}_{M_{J0} K_{J0}} (\alpha_J,\beta_J,\gamma_J)  D^{\pi_0 *} (\varphi_\pi) , \nonumber \\
  D^{Z_0 *} (\varphi_Z) &= e^{i\varphi_Z Z_0} , \\
  D^{N_0 *} (\varphi_N) &= e^{i\varphi_N N_0} , \\
  D^{A_0 *} (\varphi_A) &= e^{i\varphi_A A_0} , \\
  D^{J_0 *}_{M_{J0} K_{J0}} (\alpha_J,\beta_J,\gamma_J) &= e^{i\alpha_{J} M_{J0}} d^{J_0}_{M_{J0} K_{J0}}(\beta_J) e^{i\gamma_{J} K_{J0}} \\
  D^{\pi_0 *} (\varphi_\pi) &= e^{i \varphi_\pi (\pi_0 - 1)/2} , 
\end{align}
\end{subequations}
where $d^{J_0}_{M_{J0} K_{J0}}$ is a (real) Wigner $d$-function \cite{Varshalovich88a}.
The rotation operators have the expressions
\begin{subequations}
\begin{align}
  R(\Omega) &\equiv  R^{Z} (\varphi_Z)  R^{N} (\varphi_N)  R^{A} (\varphi_A) \\ 
            &\phantom{==} R^{J} (\alpha_J,\beta_J,\gamma_J)  R^{\pi} (\varphi_\pi) , \nonumber \\
  R^{Z} (\varphi_Z) &= e^{-i\varphi_Z Z} , \\
  R^{N} (\varphi_N) &= e^{-i\varphi_N N} , \\
  R^{A} (\varphi_A) &= e^{-i\varphi_A A} , \\
  R^{J} (\alpha_J,\beta_J,\gamma_J) &=  e^{-i \alpha_J J_z} e^{-i \beta_J J_y} e^{-i \gamma_J J_z} , \\
  R^{\pi} (\varphi_\pi) &= e^{-i \varphi_\pi (\Pi - A)/2} , 
\end{align}
\end{subequations}
where $Z$, $N$, $A$, $J_z$, $J_y$ and $\Pi$ are the one-body operators associated with the number of protons, the number of neutrons, the number of nucleons, the $z$-compondent of the angular momentum, the $y$-component of the angular momentum and the parity, respectively.
Within the SHO basis, these operators take the simple form
\begin{subequations}
\begin{align}
 Z &= \sum_p \delta_{m_{t_p} -1/2} \, c^\dagger_p c_p ,\\
 N &= \sum_p \delta_{m_{t_p} +1/2} \, c^\dagger_p c_p ,\\
 A &= \sum_p c^\dagger_p c_p ,\\
 J_z &= \sum_p m_{j_p} \, c^\dagger_p c_p ,\\
 J_y &= \sum_{pq}  \frac{i}{2} \left[ \sqrt{j_p (j_p + 1) - m_{j_p} ( m_{j_p} + 1}) \, \delta_{m_{j_q} m_{j_p}+1} \right. \\
 & -  \left. \sqrt{j_p (j_p + 1) - m_{j_p} ( m_{j_p} - 1}) \, \delta_{m_{j_q} m_{j_p}-1} \right] 
 \delta_{\hat{p} \hat{q}} \, c^\dagger_p c_q , \\
 \Pi & = \sum_p (-1)^{l_p} \, c^\dagger_p c_p ,
\end{align}
\end{subequations}
where we used the notations introduced in Ref.~\cite{Bally21b} and, as there is no risk of ambiguity, we used the index $q$ to label to the single-particle states.

Concerning the particle-number and angular-momentum projections, the integrals are discretized following the prescriptions described in Ref.~\cite{Bally21a}. In addition, given the high computational cost of multi-dimensional symmetry restoration, we employ an hybrid OpenMP+MPI parallelization scheme (see \ref{sec:para} and Ref.~\cite{Bally21b} for more details) to carry out large-scale computations.

\subsubsection{Rotated reference states}

We can also define the rotated Bogoliubov quasiparticle state
\begin{equation}
 \ket{\Phi_a(\Omega)} = R(\Omega) \ket{\Phi_a} ,          
\end{equation}
whose Bogoliubov matrices $(U_a (\Omega),V_a (\Omega))$ can be expressed in terms of the ones $(U_a,V_a)$ of $\ket{\Phi_a}$ through
\begin{subequations}
\begin{align}
 U_a (\Omega) &= \mathcal{R}(\Omega) U_a , \\
 V_a (\Omega) &= \mathcal{R}^*(\Omega) V_a , 
\end{align}
\end{subequations}
where $\mathcal{R}(\Omega)$ is the rotation matrix 
\begin{subequations}
\begin{align}
  \mathcal{R}(\Omega) &\equiv \mathcal{R}^{Z} (\varphi_Z)  \mathcal{R}^{N} (\varphi_N)  \mathcal{R}^{A} (\varphi_A) \\ &\phantom{==} \mathcal{R}^{J} (\alpha_J,\beta_J,\gamma_J)  \mathcal{R}^{\pi} (\varphi_\pi) , \nonumber \\
  \mathcal{R}^{Z} (\varphi_Z)_{pq} &= e^{-i \varphi_Z \delta_{m_p -1/2}} \delta_{pq} , \\
  \mathcal{R}^{N} (\varphi_N)_{pq} &= e^{-i \varphi_N \delta_{m_p +1/2}} \delta_{pq} , \\
  \mathcal{R}^{A} (\varphi_A)_{pq} &= e^{-i \varphi_A} \delta_{pq} , \\
  \mathcal{R}^{J} (\alpha_J,\beta_J,\gamma_J)_{pq} &= e^{-i \alpha_J m_{j_p}} d^{j_p}_{m_{j_p} m_{j_q}} (\beta_J) 
      e^{-i \gamma_J m_{j_q}} \delta_{\hat{p} \hat{q}} , \\
  \mathcal{R}^{\pi}_{pq} (\varphi_\pi) &= e^{-i \varphi_\pi ( (-1)^{l_p} - 1)/2} \delta_{pq} .
\end{align}
\end{subequations}

\subsubsection{Matrix elements}

The computation of the matrix elements $\elma{\Psi^{\lambda' i'}_{{\Sigma \epsilon'}}}{O}{\Psi^{\lambda i}_{\Sigma \epsilon}}$ between the correlated PGCM states amounts, ultimately, to the computation of matrix elements of the form $\elma{\Phi_a}{O}{\Phi_b (\Omega)}$ between the reference states of the set $\Sigma$. 
To calculate the latter, we make use of the Generalized Wick Theorem of Balian and Brézin \cite{Balian69a}, which is applicable under the assumption that $\scal{\Phi_a}{\Phi_b (\Omega)} \neq 0$.\footnote{Note that, in some cases, reference states that are orthogonal before rotation, i.e., $\scal{\Phi_a}{\Phi_b(0)} = 0$, for example for symmetry reasons, may become non-orthogonal for non-zero values of $\Omega$ such that they can still be handled using a clever discretization of the integral over $\Omega$ \cite{Bally21a}.}\footnote{Also, we remark that a new formulation of the Wick Theorem, well behaved even in the case of orthogonal states, has been recently proposed \cite{Chen23a}.}  Using the definitions given above, transition one-body densities can then be calculated using the formula given in Sec. 2.3.2 of Ref.~\cite{Bally21b} taking $\ket{\Phi_L} \equiv \ket{\Phi_a}$ and $\ket{\Phi_R} \equiv \ket{\Phi_b (\Omega)}$, and from them the non-diagonal matrix elements for any operator of interest \cite{Balian69a}.

In the specific case of the norm overlap, i.e., considering the identity operator $O \equiv 1$, we make use of the Pfaffian algebra \cite{Robledo09a,Robledo11a} and more specifically of the formula derived in Ref.~\cite{Avez12a}. While the Pfaffian method has proved to be a robust method to compute complex norm overlaps without phase ambiguity, other techniques have been developed in recent years that could be used in the future \cite{Bally18a,Mizusaki18a,Porro22a,Neergard23a}.

Similarly to the code \TAURUSvap, we consider here only a Hamiltonian that contain terms up to two-body operators. This implies that the effect of high-body terms can only be taken into account through the normal-ordered two-body approximation \cite{Roth12a,Gebrerufael16a} or a similar operator rank-reduction scheme \cite{Frosini21a}.

In addition to the overlap and the energy, the operators presently evaluated include:
\begin{itemize}
  \item the basic operators related to the quantum numbers, i.e., the number of protons ($Z$), neutrons ($N$), nucleons ($A$), the parity ($\pi$), the total angular momentum ($J$) and the isospin ($T$).
  \item the root-mean-square (rms) proton ($r_p$), neutron ($r_n$), matter ($r_a$) and charge ($r_{ch}$) radii. In this case, the codes can consider either the plain one-body operators or the two-body center-of-mass-corrected operators as defined in Ref.~\cite{Cipollone15a}. The rms charge radius takes into account the corrections for the finite size of nucleons \cite{Cipollone15a}, the Darvin-Foldy relativistic correction \cite{Friar97a} and the spin-orbit correction \cite{Reinhard21a,Reinhard23a}.
  \item the reduced probabilities $B(T\lambda)$ for the electromagnetic transitions $T \lambda \equiv E1, E2, E3, M1, M2$, using their textbook one-body expressions \cite{RS80a,BallyPHD}.
  \item the electric quadrupole ($Q_s$) and magnetic dipole ($\mu$) moments, using their textbook one-body expressions \cite{RS80a,BallyPHD}.
\end{itemize}

\subsubsection{Generalized eigenvalue problem}

Injecting the anstaz of Eq.~\eqref{eq:ansatz} into the variational equation \eqref{eq:var}, we obtain, for each symmetry block $\lambda$, a GEP of the form
\begin{equation}
\label{eq:gepdef}
  H^{\lambda \Sigma} f = e N^{\lambda \Sigma} f ,
\end{equation}
where $H^{\lambda \Sigma}$ and $N^{\lambda \Sigma}$ are the Hamiltonian and norm matrices, respectively, with matrix elements 
\begin{subequations}
\begin{align}
 H^{\lambda \Sigma}_{(ai)(bj)} &= \elma{\Phi_a}{H P^\lambda_{ij}}{\Phi_b} , \\
 N^{\lambda \Sigma}_{(ai)(bj)} &= \elma{\Phi_a}{P^\lambda_{ij}}{\Phi_b} . 
\end{align}
\end{subequations}
Here, we collected the indices for the number of reference states and the dimension of the irrep into a super index: $(ai) \in \llbracket 1, n_\Sigma \rrbracket \times \llbracket 1, d_\lambda \rrbracket$.\footnote{In particular, this implies that we do not perform any partial pre-diaganolization within each symmetry block as done, for example, with the so-called $K$-mixing when projecting onto a good angular momentum \cite{Bally21a}.} The solution of the GEP are the generalized eigenvalues $e$ and generalized eigenvectors $f$. 
At maximum, there are $n_\Sigma d_\lambda $ such solutions. In practice, however, 
we apply cutoffs when solving the GEP to avoid linear redundancies and numerically problematic states such that the actual number of solutions is often smaller than this maximum.

More precisely, considering the quantum numbers $(\lambda;j) \equiv (Z_0,N_0,A_0,J_0,\pi_0;M_{J0})$, before forming the Hamiltonian and norm matrices, we start by discarding any projected component $P^\lambda_{jj}\ket{\Phi_a}$ that is such that
\begin{subequations}
\begin{align}
 \elma{\Phi_a}{P^\lambda_{jj}}{\Phi_a} &< \varepsilon_1 , \\
 \elma{\Phi_a}{ H P^\lambda_{jj}}{\Phi_a}_n  &> \varepsilon_H , \\
 \left\vert \elma{\Phi_a}{ Z P^\lambda_{jj}}{\Phi_a}_n - Z_0 \right\vert &> \varepsilon_A , \\
 \left\vert \elma{\Phi_a}{ N P^\lambda_{jj}}{\Phi_a}_n - N_0 \right\vert &> \varepsilon_A , \\
 \left\vert \elma{\Phi_a}{ A P^\lambda_{jj}}{\Phi_a}_n - A_0 \right\vert &> \varepsilon_A , \\
 \left\vert \elma{\Phi_a}{ J_z P^\lambda_{jj}}{\Phi_a}_n - M_{J0}\right\vert &> \varepsilon_J , \\
 \left\vert \frac12 \left( -1 + \sqrt{1 + 4 \elma{\Phi_a}{ J^2 P^\lambda_{jj}}{\Phi_a}_n} \right) - J_{0}\right\vert &> \varepsilon_J , 
\end{align}
\end{subequations}
where $\varepsilon_1$, $\varepsilon_H$, $\varepsilon_A$, $\varepsilon_J$ ared numerical values selected when running the calculation and we consider normalized matrix elements
\begin{equation}
  \elma{\Phi_a}{ O P^\lambda_{jj}}{\Phi_a}_n \equiv \frac{\elma{\Phi_a}{ O P^\lambda_{jj}}{\Phi_a}}{\elma{\Phi_a}{P^\lambda_{jj}}{\Phi_a}} \, .
\end{equation} 

Then, we diagonalize the norm matrix 
\begin{equation}
\label{eq:gepd1}
 D^T N^{\lambda \Sigma} D = \text{diag}(n_\epsilon) ,
\end{equation}
and construct the normalized eigenbasis discarding any norm eigenvalue that is smaller than a value value, defined relatively to the maximum eigenvalue, 
\begin{equation}
\begin{split}
 \bar{D}_{pq} = \left\{ \begin{array}{ll} {D}_{pq}/\sqrt{n_q} &\quad \text{if } n_q \ge \max(n_\epsilon)\varepsilon_n ,   \\ 0 &\quad \text{otherwise} . \end{array}    \right.    
\end{split}
\end{equation}
The value of the cutoff $\varepsilon_n$ is also entered in input when launching the calculation.

To obtain the generalized eigenvalues of the problem, we diagonalize the Hamiltonian matrix in this basis
\begin{equation}
\label{eq:gepd2}
 C^T \left( \bar{D}^T H^{\lambda \Sigma} \bar{D} \right) C = \text{diag}(e_\epsilon) .
\end{equation}
They represent the energies $E_{\Sigma \epsilon}^{\lambda} = e_\epsilon$ of the correlated wave functions introduced in Eq.~\eqref{eq:ansatz}. The weights $f^{\lambda i}_{\Sigma a \epsilon}$ are obtained by computing the generalized eigenvectors 
\begin{align}
 f^{\lambda i}_{\Sigma a \epsilon} &= (\bar{D} C)_{(ai) \epsilon} .
\end{align}

Finally, we indicate that the diagonalizations in Eqs.~\eqref{eq:gepd1} and \eqref{eq:gepd2} are performed using the routine \textsf{DSYEV} of the library \textsf{LAPACK}. We mention in passing that the library also offers the possibility to directly solve Eq.~\eqref{eq:gepdef} using the routine \textsf{DGGEV}, which is based on the QZ algorithm. While this latter method is sometimes useful, it is not as flexible as the approach discussed here as it does not allow us for removing numerically unstable states in the intermediate step represented by Eq.~\eqref{eq:gepd1}.

%
%
\section{Example: low-energy structure of $^{24}$Mg with a chiral interaction}
\label{sec:example}

\subsection{Motivations and setting}

As example of realistic application, we study the low-energy structure of $^{24}$Mg, which has often been used in the past to benchmark PGCM calculations based on phenomenological energy density functionals \cite{Bender08a,Rodriguez10a,Yao10a}. Being located in the middle of the sd-shell, the isotope of  $^{24}$Mg exhibits a triaxial structure with well-identified rotational bands. It is, therefore, a good test case for PGCM calculations that include the full three-dimensional angular-momentum projection. 

In the present study, we perform no-core PGCM calculations based on Bogoliubov quasiparticle states optimized by varying the particle-number projected energy using the code \TAURUSvap. We take as generator coordinates the triaxial deformations in the first sextant of the $(\beta, \gamma)$ plane \cite{RS80a} and perform the configuration mixing of the reference states projected onto all the appropriate quantum numbers.
Concerning the nuclear Hamiltonian, we employ the $\chi$EFT-based interaction known as ``EM1.8/2.0'' \cite{Hebeler11a}, which has been used extensively in \emph{ab initio} calculations over the last decade \cite{Hergert20a,Stroberg19a,Tichai23a,Yao20a}. The full interaction is reduced to an effective two-body operator using the scheme presented in Ref.~\cite{Frosini21a}.

To explore triaxial deformations, we employ a parallelogram mesh in $(\beta,\gamma)$, as described in Ref.~\cite{Rodriguez10a}, and consider only the first of sextant of the $(\beta, \gamma)$ plane.\footnote{For reference states that break the time-reversal invariance, in principle, it should be necessary to explore the three non-equivalent sextants of the triaxial plane \cite{Schunck10a}. We do not expect this approximation to greatly affect our results.} To be more precise, along the axial axis, our choice corresponds to a mesh starting at $\beta = 0$ and with a spacing $\Delta \beta = 0.05$.

Concerning the angular-momentum projection, we discretize the triple integral over Euler angles $(\alpha_J,\beta_J,\gamma_J) \in [0,2\pi] \times [0,\pi] \times [0, 2\pi]$ using $32 \times 20 \times 32$ points. For the particle-number projection, we discretize the double integral over gauge angles $(\varphi_Z,\varphi_N)$ using $5 \times 5$ points in the interval $[0,\pi] \times [0,\pi]$ when dealing with reference states that do not include proton-neutron mixing, and $7 \times 7$ points in the interval $[0,2\pi] \times [0,2\pi]$ when dealing with more general reference states that do.
These values are sufficient to obtain excellent convergence of the discretized projection operators.

Finally, during the configuration mixing, we set the cutoffs to the values: $\varepsilon_1 = 10^{-5}$, $\varepsilon_H = -10^{+2}$, $\varepsilon_A = 10^{-3}$, $\varepsilon_J = 10^{-3}$ and $\varepsilon_n = 10^{-6}$. In particular, we checked that by varying simultaneously the values of $\varepsilon_1 \in \left\{ 10^{-4}, 10^{-5}, 10^{-6}\right\}$ and $\varepsilon_n \in \left\{ 10^{-5}, 10^{-6}, 10^{-7}, 10^{-8} \right\}$, the results for the different observables after configuration mixing were stable. More precisely, while some small variations can be observed when changing the values of the cutoffs, they are limited and almost always much smaller than the differences found between the various configuration mixing discussed below.

\subsection{Projected energy surface for $J_\epsilon^\pi = 0^+_1$}

\begin{figure}[t!]
\centering  
  \includegraphics[width=1.00\columnwidth]{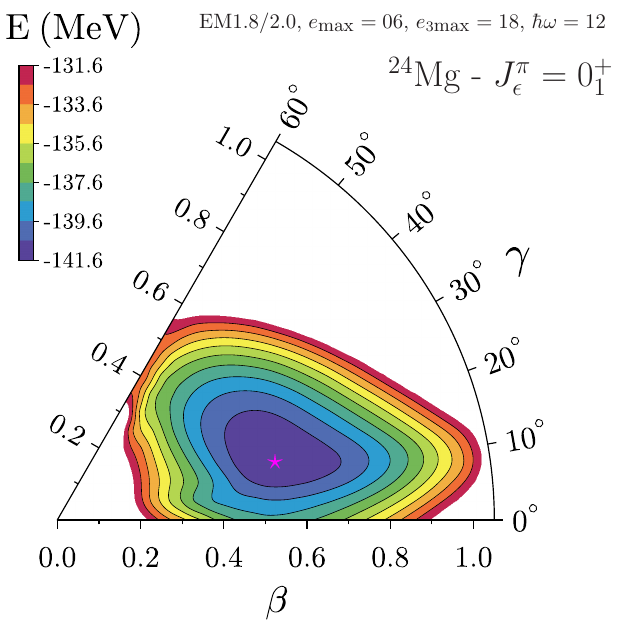} 
\caption{
\label{fig:surfacce_mg24}
(Color online)
Projected energy surface for $^{24}$Mg and $J_\epsilon^\pi = 0^+_1$ in the first sextant of the $(\beta, \gamma)$ plane. The black lines are separated by 1 MeV, starting from the minimum represented by a magenta star, which is located at a deformation of $\beta \approx 0.54$ and $\gamma \approx 15^\circ$.
The calculations were performed with the interaction EM1.8/2.0 in a model space with $e_{\text{max}} = 6$, $e_{\text{3max}} = 18$ and $\hbar \omega = 12$
}
\end{figure}

We start by considering reference states that are time-reversal as well as parity invariant and can be written as a direct product of separate proton and neutron wave functions, i.e., without proton-neutron mixing, 
represented in a model space with $e_{\text{max}} = 6$, $e_{\text{3max}} = 18$ and $\hbar \omega = 12$. While this basis is of small size, in the past it proved to be sufficient to converge calculations of light systems with mass numbers similar to the one of $^{24}$Mg \cite{Frosini22b,Frosini22c}.

In Fig.~\ref{fig:surfacce_mg24}, we display the $J_\epsilon^\pi = 0^+_1$ projected total energy surface,\footnote{Here, we mention that we plot the projected energy at the average deformation of the underlying Bogoliubov reference state.} up to 10 MeV above the minimum.
As expected for this nucleus, we obtain an energy surface centered around a triaxial minimum, with a deformation of $\beta \approx 0.54$ and $\gamma \approx 15^\circ$, and stretching towards larger values of $\beta$ at small values of $\gamma$. This is in agreement with previous multi-reference energy density functional calculations (MREDF) \cite{Bender08a,Rodriguez10a} although, in the present case, the energy surface is more rigid against $\beta$-deformation. We also remark that the minimum is underbound by approximately 57 MeV compared to the experimental energy,\footnote{While the correct comparison should be with the exact theoretical ground state of $^{24}$Mg obtained with the same Hamiltonian, we expect the energy of the latter to be somewhat close to the experimental value.} which is due to the missing weak correlations in our approach \cite{Tichai18a,Frosini22a,Frosini22b,Frosini22c,Duguet23a}. 

\subsection{Selection scheme and convergence study}

To choose the reference states to be included in the configuration mixing, we start by sorting them according to the energy of their $J_\epsilon^\pi = 0^+_1$ projected state and then, we select all the Bogoliubov states that give a projected energy lower than a given threshold, $E_{\text{thr}}$, above the projected minimum with same quantum numbers. A similar strategy was used, for example, in Refs.~\cite{Bally22b,Bally23a}.

While not perfect, this method has the advantage of being simple, easily reproducible, and physically sound for low-lying collective states built on top of the $0^+_1$ ground state, as the structure (e.g. energy surfaces) of the former resembles the one of the latter.
Nonetheless, we remark that more sophisticated strategies exist
that take into account the contribution of the reference states to the configuration mixing \cite{Otsuka01a,Dao22a,Matsumoto23a}. Most of them, however, have been so far used only in small model spaces, and it is not clear yet if they are also tractable in large-scale no-core PGCM calculations.

Another important step in our calculations is to determine how converged are the PGCM results with respect to a change of the size of the basis ($e_{\text{max}}$), the deformation mesh ($\Delta \beta$) or the energy threshold ($E_{\text{thr}}$) employed to select the states. In that order, we perform several different calculations varying these parameters. More precisely, we consider the following computations:
\begin{itemize}
  \item $\text{PGCM}_1$ with $e_{\text{max}} = 6$, $\Delta \beta = 0.05$, $E_{\text{thr}} = 5$ MeV. The mixing set contains 90 reference states.
  \item $\text{PGCM}_2$ with $e_{\text{max}} = 6$, $\Delta \beta = 0.10$, $E_{\text{thr}} = 5$ MeV. The mixing set contains 24 reference states.
  \item $\text{PGCM}_3$ with $e_{\text{max}} = 6$, $\Delta \beta = 0.10$, $E_{\text{thr}} = 10$ MeV. The mixing set contains 39 reference states.
  \item $\text{PGCM}_4$ with $e_{\text{max}} = 8$, $\Delta \beta = 0.10$, $E_{\text{thr}} = 10$ MeV. The mixing set contains 33 reference states.
\end{itemize}
Note that, in all the above calculations, we fix $e_{\text{3max}} = 18$, $\hbar \omega = 12$ and consider reference states that are time-reversal invariant, parity invariant and do not include proton-neutron mixing.

On the one hand, the PGCM is a variational approach and, thus, will ultimately converge to the exact results if the mixing set is enriched enough. On the other hand, here we do not include explicitly in the set the large number of particle-hole configurations that would bring the necessary weak correlations to reach the exact eigenenergies. For these reasons, there is little hope to obtain the convergence of absolute energies in our calculations as we increase the number of reference states included. On the contrary, we will observe a slow decrease of the absolute energies as we increase the size of the mixing set and the variational principle finds (very small) additional contributions to the energy. With that in mind, we will mainly focus our analysis on the convergence of relative quantities such as the excitation energies.

\begin{figure*}[t]
\centering
\includegraphics[width=.800\linewidth]{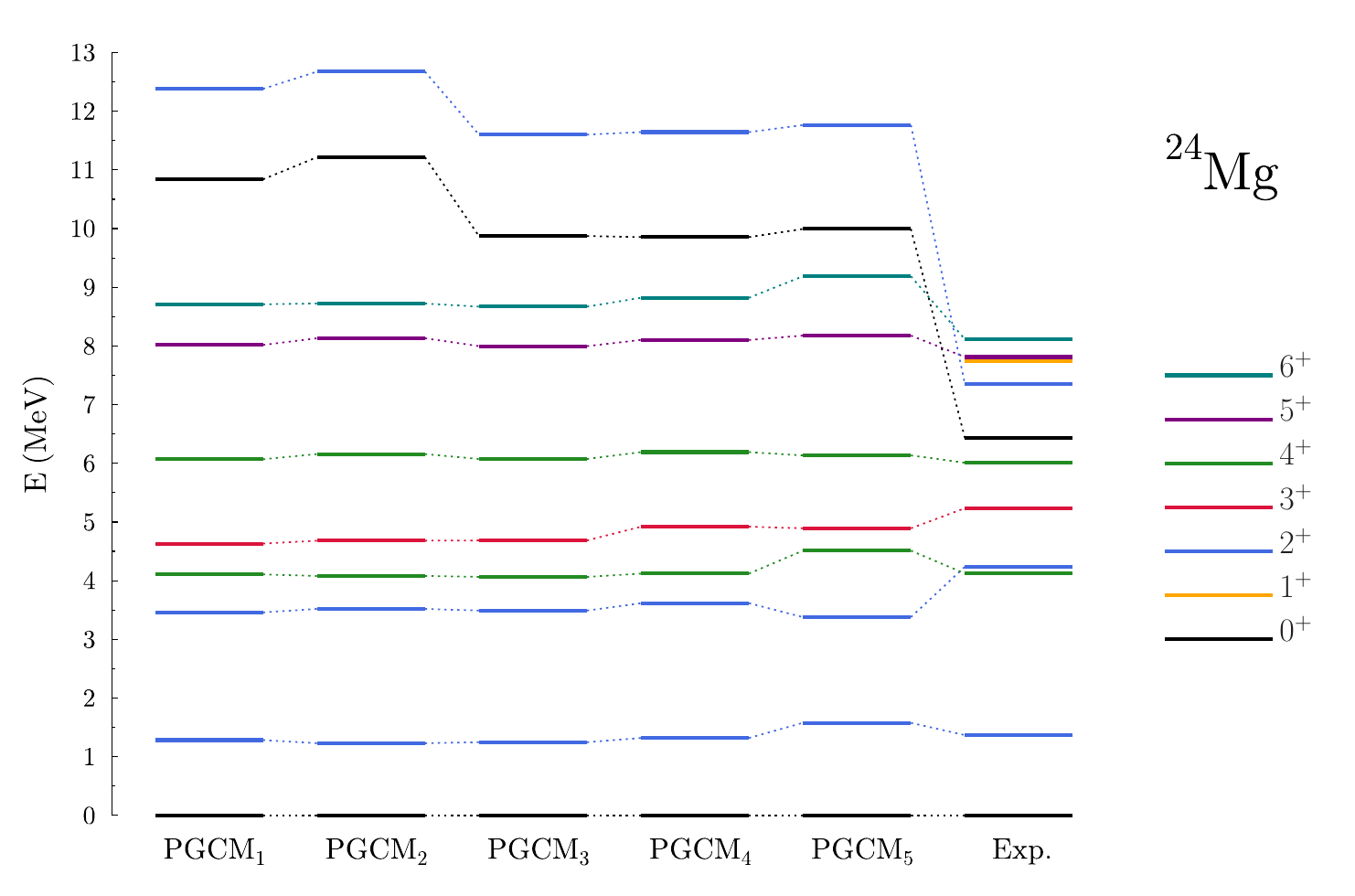}
\caption{
Low-energy spectrum for positive parity states of $^{24}$Mg up to the $6^+_1$ experimental state.
For theoretical results, states that correspond to one of the low-lying experimental states considered are plotted even if their energy is above the one of the $6^+_1$ experimental state.
Experimental data are taken from Ref.~\cite{Shamsuzzoha22a}
}
\label{fig:spectrum_conv_mg24}
\end{figure*}

\begin{table*}[h!]
\centering
\begin{tabular}{cccccccc}
Quantity & $\text{PGCM}_1$ & $\text{PGCM}_2$ & $\text{PGCM}_3$ & $\text{PGCM}_4$ & $\Delta_{1234}$ & $\text{PGCM}_5$ & Experiment \\
\hline
$E(2^+_1)$ &  1.284 &  1.231 &  1.248 &  1.320 & 0.089 & 1.578 & 1.369 \\[0.08cm]
$E(4^+_1)$ &  4.107 &  4.081 &  4.065 &  4.121 & 0.056 & 4.512 & 4.123 \\[0.08cm]
$E(2^+_2)$ &  3.460 &  3.519 &  3.491 &  3.617 & 0.157 & 3.379 & 4.238 \\[0.08cm]
$E(3^+_1)$ &  4.632 &  4.682 &  4.686 &  4.920 & 0.288 & 4.893 & 5.235 \\[0.08cm]
$E(4^+_2)$ &  6.069 &  6.158 &  6.074 &  6.191 & 0.122 & 6.135 & 6.010 \\[0.08cm]
$E(0^+_2)$ & 10.839 & 11.216 &  9.873 &  9.855 & 1.361 & 9.994 & 6.432 \\[0.08cm]
$E(2^+_3)$ & 12.381 & 12.674 & 11.601 & 11.644 & 1.073 & 11.764 & 7.349 \\[0.08cm]
$E(1^+_1)$ &        &        &        &        & & 41.762 & 7.748 \\[0.08cm]
$E(5^+_1)$ &  8.017 &  8.131 &  7.997 &  8.102 & 0.134 &  8.177 & 7.812 \\[0.08cm]
$E(6^+_1)$ &  8.711 &  8.723 &  8.670 &  8.821 & 0.151 &  9.188 & 8.113 \\[0.08cm]
$r_\text{rms}(0^+_1)$ & 3.000 & 2.999 & 3.001 & 3.000 & 0.002 & 2.999 & 3.057(2) \\[0.08cm]
$\mu(2^+_1)$ & +1.03 & +1.03 & +1.02 & +1.04 & 0.02 & +1.03 & +1.08(3) \\[0.08cm]
$\mu(2^+_2)$ & +1.06 & +1.06 & +1.05 & +1.06 & 0.01 & +1.06 & +1.3(4) \\[0.08cm]
$\mu(4^+_1)$ & +2.06 & +2.05 & +2.05 & +2.06 & 0.01 & +2.06 & +1.7(12) \\[0.08cm]
$\mu(4^+_2)$ & +2.06 & +2.06 & +2.05 & +2.07 & 0.02 & +2.07 & +2.1(16) \\[0.08cm]
$Q_s(2^+_1)$ & -17.2 & -17.2 & -17.2 & -17.0 & 0.2 & -15.4 & -29(3) \cite{Stone21a} \\[0.08cm]
             & & & & & & & -18(2) \cite{Spear81a} \\[0.08cm]
$Q_s(2^+_2)$ & +17.6 & +17.7 & +17.6 & +17.4 & 0.3 & +15.5 & \\[0.08cm]
$Q_s(3^+_1)$ & -0.1  & -0.1  & -0.1  & -0.1  & 0.0 & 0.0 & \\[0.08cm]
$Q_s(4^+_1)$ & -21.1 & -21.0 & -21.3 & -21.5 & 0.4 & -19.8 & \\[0.08cm]
$B(E2:2^+_1 \rightarrow 0^+_1$) & 75 & 75 & 75 & 76 & 1 & 76 & 87(2) \\[0.08cm]
$B(E2:4^+_1 \rightarrow 2^+_1$) & 103 & 103 & 102 & 103 & 1 & 108 & 147(+14-12) \\[0.08cm]
\hline
\end{tabular}
\caption{Spectroscopic quantities for some low-lying states $(J^\pi_\epsilon)$ of $^{24}$Mg: excitation energy $E$ (MeV),  root-mean-square charge radius $r_\text{rms}$ (fm), magnetic dipole moments $\mu$ ($\mu_N$), spectroscopic quadrupole moments $Q_s$ ($e\, \text{fm}^2$) and reduced transition probabilities $B(E2)$ ($e^2\, \text{fm}^4$).
The column $\Delta_{1234}$ is defined as the absolute value of the largest difference between the calculations $\text{PGCM}_k$, $k=1,2,3$ or 4, for each observable. Experimental uncertainties are relative to the last digit written.
Experimental data are taken from \cite{Angeli13a,Stone20a,Stone21a,Spear81a,Shamsuzzoha22a}}
\label{tab:quantities_conv_mg24}
\end{table*}

In Fig.~\ref{fig:spectrum_conv_mg24}, we show the energy spectra obtained from the four different PGCM calculations listed above for positive parity states associated with a low-lying experimental level up to the $6^+_1$ (experimental) state. The exact energy of the states can be found in Table~\ref{tab:quantities_conv_mg24}.
Obviously, the most striking observation is that the four calculations give extremely similar results for all the states up to the $6^+_1$ state located approximately at 8.75 MeV in our computations. To better quantify the agreement between the four realizations, in the column $\Delta_{1234}$ of Table~\ref{tab:quantities_conv_mg24}, we give the absolute value of the largest difference obtained between two separate calculations.
In most cases, the different calculations agree on the value of the excitation energy of a given state within a range of about 150 keV. The worst agreement is observed for the $3^+_1$ states with a difference of 288 keV between $\text{PGCM}_1$ and $\text{PGCM}_4$, which is still quite satisfying given the fact these are the two most dissimilar PGCM calculations considered here, i.e., all three parameters are different.
Larger differences of the order of 1 MeV appear for the higher-lying $0^+_2$ and $2^+_3$ states.

In Table~\ref{tab:quantities_conv_mg24}, we also report, for some of the low-lying states, other spectroscopic quantities\footnote{In the following, $\mu_N$ is the nuclear magneton whereas $e$ is the elementary charge.} such as the
rms charge radius $r_\text{rms}$ (fm), magnetic dipole moments $\mu$ ($\mu_N$), spectroscopic quadrupole moments $Q_s$ ($e\, \text{fm}^2$) and reduced transition probabilities $B(E2)$ ($e^2\, \text{fm}^4$). Here, we precise that we employ the bare values of the electric charge and $g$-factors for protons and neutrons.
As can be seen, the agreement between the various PGCM calculations is even better than for the energies. Indeed, for all observables considered here, the discrepancy on the computed value is, at most, of the order of a few percents. 

From these data, we can draw the following conclusions:
\begin{itemize}
  \item the similarity between $\text{PGCM}_1$ and $\text{PGCM}_2$ suggests that we have obtained a good convergence as function of the spacing $\Delta \beta$.\footnote{Also, it shows that we can reproduce the $\text{PGCM}_1$ results based on 90 reference states with the much smaller $\text{PGCM}_2$ that only makes use of 24 reference states. This is particularly interesting as, given the scaling of the method, the computational cost necessary to perform $\text{PGCM}_1$ is roughly 14 times larger than the one required to perform $\text{PGCM}_2$.}
  \item the similarity between $\text{PGCM}_2$ and $\text{PGCM}_3$ suggests that the results are relatively insensitive to our choice for the value of $E_\text{thr}$, except for the $0^+_2$ and $2^+_3$ states, which are possibly not fully converged even for $E_\text{thr} = 10$ MeV.
  \item the similarity between $\text{PGCM}_3$ and $\text{PGCM}_4$ suggests a good convergence as a function of $e_{\text{max}}$.
\end{itemize}
To sum up, this limited convergence study seems to indicate that our calculations
are well converged with respect to the parameters $e_\text{max}$, $\Delta \beta$ and $E_\text{thr}$. 
It would be certainly desirable to carry out a more thorough analysis\footnote{For example, a more complete convergence study of \emph{ab initio} PGCM calculations was realized in Ref.~\cite{PorroPHD}.} in the future, e.g., considering the simultaneous variation of all the parameters, but this falls outside the scope of the present example calculation.\footnote{Note that it would be also necessary to estimate the convergence as a function of the truncation order in the EFT.}

\subsection{More general reference states}

To probe the effects of exploring additional degrees of freedom in the configuration mixing, we perform another calculation, $\text{PGCM}_5$, based on real general Bogoliubov quasiparticle states. Indeed, these states allow for the supplementary 
\begin{enumerate}
  \item breaking of time-reversal invariance, which naturally leads to the inclusion of $K=1$ projected states in the mixing.
  \item spontaneous breaking of parity invariance, which is related to the development of non-vanishing expectation values for odd-rank multipole operators.  
  \item mixing of protons and neutrons, which permits to take into account both $(T=0,J=1)$ and $(T=1,J=0)$ pairing channels, a possible important feature in the case of the $N=Z$ nucleus $^{24}$Mg. 
\end{enumerate}
The calculations are performed fixing the parameters to the values: $e_{\text{max}} = 6$, $e_{\text{3max}} = 18$, $\hbar \omega = 12$, $\Delta \beta = 0.10$ and $E_{\text{thr}} = 10$ MeV. The mixing set contains 38 reference states. 

The results of the realization $\text{PGCM}_5$ are reported in Figure~\ref{fig:spectrum_conv_mg24} and Table~\ref{tab:quantities_conv_mg24}.
First, we notice that the level change observed by going from the calculations $\text{PGCM}_{1-4}$ to $\text{PGCM}_5$ is larger than the variations observed among the calculations $\text{PGCM}_{1-4}$ themselves described in the previous section. This statement particularly applies to the excitation energies and the electric quadrupole moments.
That being said, the overall picture obtained from the computation $\text{PGCM}_5$ is still very similar to the one obtained from the previous calculations. It seems, therefore, that the new degrees of freedom described above do not have a strong influence on the final results.
Although, and anticipating a little bit the discussion of Sec.~\ref{sec:compare}, we remark that the calculation $\text{PGCM}_5$ performs more poorly than the other calculations in reproducing experimental data.

A limitation of our approach is that, even though we authorize the reference states to explore additional degrees of freedom, we do not enforce it through a constrained minimization procedure during the generation of the Bogoliubov product states.
Therefore, the effective exploration of these modes depends in practice on what is energetically favored by the variation after particle-number projection (VAPNP) minimization. For example, we observe that at the bottom of the energy surface, our variational procedure mostly generates reference states that have, in good approximation, a positive parity after particle-number projection. A similar pattern appears for the pairing content of the reference states with certain regions of our triaxial surface favoring specific components among the six possibilities available in the $(T=0,J=1)$ and $(T=1,J=0)$ channels.

This shortcoming is particularly problematic for soft modes, such as octupole deformations \cite{Robledo11b,Robledo15a}, because the minimization and subsequent configuration mixing and symmetry projections may favor radically different reference states. Ultimately, this implies that we should perform more sophisticated calculations that explicitly include  these degrees of freedom in the configuration mixing through the use of reference states generated by performing adequate constrained HFB/VAPNP calculations. 

\subsection{Comparison with experimental data}
\label{sec:compare}

To better appreciate the quality of our results, we are now interested in comparing them to known experimental data \cite{Angeli13a,Stone20a,Stone21a,Spear81a,Shamsuzzoha22a}.
For the sake of clarity, we will designate calculation $\text{PGCM}_4$, which corresponds to the computation in the largest basis with $e_{\text{max}} = 8$, as reference for theoretical results. 

\begin{figure}[t]
\centering
\includegraphics[width=1.00\linewidth]{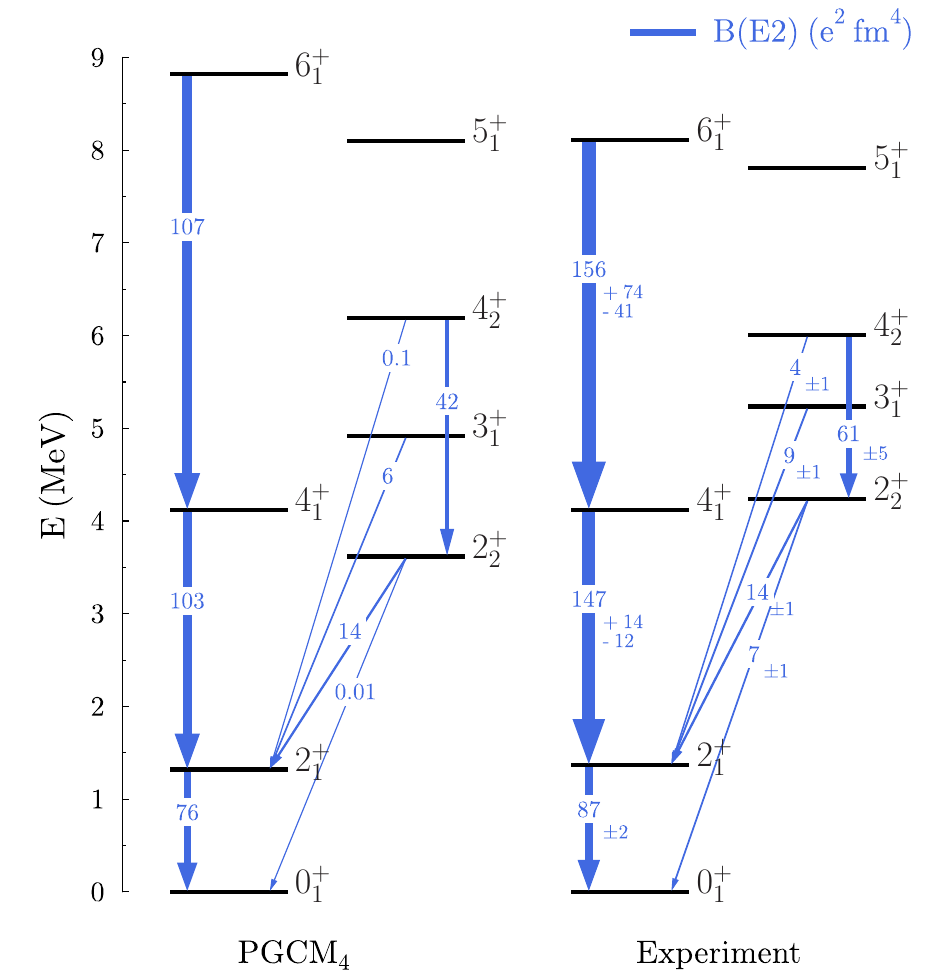}
\caption{
Ground-state and first excited rotational bands for $^{24}$Mg. Blue arrows and numbers represent reduced transition probabilities, $B(E2)$, given in unit of $e^2 \text{fm}^4$.
Experimental data are taken from Ref.~\cite{Shamsuzzoha22a}
}
\label{fig:rotband_mg24}
\end{figure}

Experimental data are reported in Fig.~\ref{fig:spectrum_conv_mg24} and  Table~\ref{tab:quantities_conv_mg24}. In addition, for a better visual representation, we also plot in Fig.~\ref{fig:rotband_mg24} the direct comparison between empirical data and $\text{PGCM}_4$ for the ground-state and first excited rotational bands. 

First of all, we remark that the global picture of the energy spectrum is relatively well described by our calculations. In particular, most of the theoretical states have an excitation energy quite close to the one of their experimental counterpart. Still, some deficiencies are apparent. First, the $2^+_2$ and $4^+_1$ states are inverted in our calculations compared to experimental results. Then, the irregular spacing between the $2^+_2$, $3^+_1$ and $4^+_2$ states belonging to the $\gamma$ band is not reproduced in our calculations. Furthermore, the computed $0^+_2$ and $2^+_3$ states are located more than 3 MeV above their experimental analogues. This defect might be due, in part, to the selection strategy employed to select the reference states that is strongly biased towards the $0^+_1$ state.
Last but not least, the low-lying $1^+_1$ state is missing in our calculations. While this is to be expected for $\text{PGCM}_{1-4}$ calculations that are based on time-reversal and parity invariant Bogoliubov quasiparticle states\footnote{Although, to be precise, this property also requires a good signature, as defined in Refs.~\cite{Dobaczewski00a,Dobaczewski00b}, which is not strictly enforced here but is still probably realized in practice up to a good approximation.} \cite{Bally14a}, this is more surprising for the calculation $\text{PGCM}_5$ that is constructed using more general reference states with no such symmetry restrictions. Actually, there exists a $1^+_1$ state in the spectrum of the calculation $\text{PGCM}_5$ but it is located at at an excitation energy of more than 40 MeV.
The good description of $J^\pi = 1^+$ states thus requires to generate the reference states through a different procedure \cite{Bofos24a}.

Similarly, the $B(E2)$ reduced transition probabilities are reasonably well described. Importantly, as can be seen on Fig.~\ref{fig:rotband_mg24}, we can clearly identify the ground-state and first excited rotational bands. Nevertheless, the strengths of several transitions are underestimated. 
Even accounting for experimental uncertainty, the intra-band transitions $4^+_1 \rightarrow 2^+_1$ and $6^+_1 \rightarrow 4^+_1$ are somewhat not collective enough. Also, the $\Delta J =2$ inter-band transitions $2^+_2 \rightarrow 0^+_1$ and $4^+_2 \rightarrow 2^+_1$ almost vanish in our calculations.
While not reported here, we want to mention that the computed $B(M1)$ reduced transition probabilities are smaller than the experimental measurements by an order of magnitude or more. This deficiency also appears in MREDF calculations \cite{Bender08a}. 

Concerning the other spectroscopic quantites reported in Table~\ref{tab:quantities_conv_mg24}, we first see that the theoretical value for the rms charge radius $r_\text{rms}(0^+_1) = 3.000$ fm is slightly too small compared to the experimental measurement of 3.057(2) fm. While the interaction EM1.8/2.0 is known to underestimate this particular observable \cite{Lapoux16a}, in our case the theoretical value is actually not small enough compared to what could be expected from previous calculations. For example, in oxygen isotopes the residuals on $r_\text{rms}(0^+_1)$ are of the order of 0.3 fm \cite{Ekstrom15a}.
The fact is that the PGCM itself tends to overestimate the rms charge radii due to the lack of weak correlations \cite{Frosini22b}. Therefore, the two effects probably partially compensate each other in such a way that it leads to a too small, but not dramatically too small, computed value for $r_\text{rms}(0^+_1)$.

The theoretical magnetic moments $\mu(2^+_1) = +1.04~\mu_N$ and $\mu(2^+_2) = +1.06~\mu_N$ agree very well with experimental data, $+1.08(3)~\mu_N$ and $+1.3(4)~\mu_N$, respectively. While the theoretical magnetic moments $\mu(4^+_1) = +2.06~\mu_N$ and $\mu(4^+_2) = 2.07~\mu_N$ are also compatible with experimental observations, $+1.7(12)~\mu_N$ and $+2.1(16)~\mu_N$, respectively, the error bars of the latter values are so large that it is not possible to draw a firm conclusion.
It is important to stress that our numerical implementation employs only the one-body part of the dipole magnetic moment but is still capable of reproducing experimental results. We acknowledge, however, that we are dealing here with the simple example of an even-even nucleus with $N=Z$. In the general case, the inclusion of two-body currents is likely necessary \cite{Miyagi23a}.

Within our approach, the electric quadrupole moment of the $2^+_1$ state has a value of $Q_s(2^+_1) = -17~e \text{fm}^2$. This is somewhat too large compared to the current recommended experimental value of $-29~e \text{fm}^2$ \cite{Stone21a}. Nevertheless, it is interesting to remark that our result is in excellent agreement with a previous adopted value of $-18(2)~e \text{fm}^2$ \cite{Fewell79a,Spear81a} as well as with the values obtained from MREDF calculations of $^{24}$Mg based on Skyrme SLy4 ($-19.4~e \text{fm}^2$) \cite{Bender08a}, Gogny D1S ($-20.8~e \text{fm}^2$) \cite{Rodriguez10a} or covariant PC-F1 functionals  \cite{Yao11a}. 

In conclusion, our calculations give a reasonable description of the low-energy spectrum of the nucleus $^{24}$Mg that is on a par with the ones obtained from advanced phenomenological MREDF calculations \cite{Bender08a,Rodriguez10a,Yao10a}.

%
%
\section{Conclusion}
\label{sec:conclu}

In this article, we presented the numerical codes \TAURUSpav \linebreak[3] and \TAURUSmix~that, combined, can be used to perform the configuration mixing of parity, particle-number and total-angular-momentum projected real general Bogoliubov quasiparticle states represented in a SHO basis. These two codes together with the previously published software \TAURUSvap~\cite{Bally21b} form the numerical suite \TAURUS.

Using the isotope of $^{24}$Mg as example, we demonstrated the utility of our software to study the structure of deformed nuclei. The calculations, based on a realistic interaction rooted in $\chi$EFT, are able to reproduce the main features of the low-energy spectrum of $^{24}$Mg. In particular, comparison of our theoretical results with experimental data shows a good agreement for most of the spectroscopic observables of interest.
This confirms the findings of Ref.~\cite{Frosini22b,Frosini22c} that, while not including explicitly weak correlations, PGCM calculations can still be used to describe relative properties of nuclei such as excitation energies or electromagnetic transitions and moments.

The next step in the development of the suite \TAURUS~will be the improvement of the numerical algorithms employed to speed-up the calculations and make them more tractable in large model spaces.
While the current versions of the codes possess an hybrid OpenMP+MPI parallel implementation that can be used to that effect, we expect that an additional layer of GPU programming could be of great advantage to tackle heavier nuclei in large model spaces.

Finally, we also envision in the future the use of (complex) general Bogoliubov quasiparticle wave functions as reference states as well as the implementation of the projection onto a good isospin \cite{Satula10a,Satula16a}.

%
%
\section{Using the code \TAURUSpav}
\label{sec:structpav}

As a preliminary warning, we would like to stress the fact that the development of a software is an evolving process and, therefore, part of the information given here may become obsolete in future versions of the code. 

\subsection{Source files and compilation}
The numerical code \TAURUSpav~is divided into different files containing the main program, the various modules, and specific mathematical routines.
The list all the files, in the order of which they should be compiled, is the following: \\
$\tt module\_constants.f90       $ \\
$\tt module\_mathmethods.f90     $ \\ 
$\tt module\_parallelization.f90 \text{ (only if \textsf{MPI})}$ \\
$\tt module\_nucleus.f90         $ \\
$\tt module\_basis.f90           $ \\
$\tt module\_hamiltonian.f90     $ \\
$\tt module\_particlenumber.f90  $ \\
$\tt module\_angularmomentum.f90 $ \\
$\tt module\_isospin.f90         $ \\
$\tt module\_multipoles.f90      $ \\
$\tt module\_radius.f90          $ \\
$\tt module\_operators.f90       $ \\
$\tt module\_wavefunctions.f90   $ \\
$\tt module\_fields.f90          $ \\
$\tt module\_projection.f90      $ \\
$\tt module\_initialization.f90  $ \\
$\tt subroutines\_pfaffian.f     $ \\
$\tt taurus\_pav.f90             $

To use the MPI implementation, it is first required, before compilation, to remove the comment flags ``\texttt{!cmpi}'' present in the Fortran files. 
This can be done easily using the command\footnote{On BSD-based systems, it is necessary to use instead $\texttt{sed -i "" "s/\textbackslash !cmpi//g" src/*f90}$}
\begin{equation*}
\texttt{sed -i "s/\textbackslash !cmpi//g" src/*f90}
\end{equation*}
while being in the main directory of the respository.\footnote{As this command changes the source files in an irreversible manner, we recommend to first create
a temporary copy of the source files before using it. Otherwise, it will be necessary to discard the changes in the local git repository before the next compilation of the code
in sequential mode (i.e.\ without MPI).}

On the GitHub repository, we provide a makefile to compile the code that takes in entry two arguments: \texttt{FC} to specify the compiler
and \texttt{TH} to specify if we want to include OpenMP threading or not. If no argument is entered, the script will use some default values.
The correct values for the argument \texttt{FC} are: \texttt{gfortran} (default), \texttt{ifort}, \texttt{mpiifort} and \texttt{mpif90}. 
On the other hand, the argument \texttt{TH} can take the values \texttt{omp} or \texttt{none} (default). 
For example, if one wants to compile \TAURUSpav~using the \texttt{gfortran} compiler without OpenMP, one has to execute the command
\begin{equation*}
\texttt{make FC=gfortran TH=none}
\end{equation*}

For completeness, we also provide in the repository a bash script that offers comparable compilation capabilities. 
More information about how to use this script or the makefile can be found in the file \texttt{README.md} present in the main directory
of the repository.

\subsection{Execution}
Once in the directory containing the executable file, the code can be run by typing the command
\begin{equation*}
\texttt{./taurus\_pav.exe < input.txt}
\end{equation*}
where \texttt{input.txt} is the STDIN file containing the input parameters. The details concerning the format of STDIN can be found in the 
file \texttt{manual\_input.pdf} present in the directory \texttt{doc}. The code also requires, in the same directory, the files containing the wave functions of the left and right reference states as well as the file defining the Hamiltonian (and model space), 
the name of which is written in the STDIN, and the various files containing its matrix elements. See the file \texttt{doc/manual\_input.pdf} for more details.
The details concerning the format of Hamiltonian files can be found in the file \texttt{manual\_hamiltonian.pdf} present in the directory \texttt{doc} of the repository associated with  \TAURUSvap~\cite{Bally21b}. 

To simplify the execution of the code, we provide the script \texttt{launch.sh} that performs all the necessary steps to run a calculation. 
To use it, go to the main directory of your copy of the repository and type the command 
\begin{equation*}
\texttt{bash launch.sh}
\end{equation*}

During its execution, the code prints various information (e.g. inputs and model space used, decomposition of reference states in terms of its different $\lambda$-components, etc.) in the STDOUT. 
We recommend to store the printing in a file, for example \texttt{output.txt}, by typing 
\begin{equation*}
\texttt{./taurus\_pav.exe < input.txt > output.txt}
\end{equation*}
or
\begin{equation*}
\texttt{bash launch.sh > ouput.txt}
\end{equation*}

Additionally, the code will produce other files containing the matrix elements of various operators to be used as inputs when running \TAURUSmix.
The names of all the files produced during a run are recalled in the STDOUT.

\subsection{Parallelization}
\label{sec:para}

The parallelization strategy used in \TAURUSpav~is almost identical to the one described in Ref.~\cite{Bally21b} for \TAURUSvap. The only difference is that the total number of angles, $N_\text{angles}$, is the product of the number of angles for each symmetry projection, i.e.,
\begin{equation}
  N_\text{angles} = N_{\varphi_Z} N_{\varphi_N} N_{\varphi_A} N_{\alpha_J} N_{\beta_J} N_{\gamma_J} N_{\varphi_\pi}, 
\end{equation}
where $N_{\varphi_Z}$, $N_{\varphi_N}$, $N_{\varphi_A}$, $N_{\phi_Z}$, $N_{\alpha_J}$ $N_{\beta_J}$, $N_{\gamma_J}$, $N_{\varphi_\pi}$ are the number of discretization points in the sums over the proton, neutron and nucleon gauge angles, the three Euler angles and the parity angle, respectively. 
To say it differently, the architecture of the code is such that there is one main outer loop regrouping all the symmetry angles and the parallelization scheme distributes the values in the loop among the different teams (see Sec.~5.4 in Ref.~\cite{Bally21b}).

\subsection{Two-body matrix elements of the Hamiltonian}
\label{sec:2BME}

The code \TAURUSpav~uses the same strategy as the one employed in \TAURUSvap~to store the two-body matrix elements of the nuclear Hamiltonian. That being said, it is important to mention that we changed the order of the indices $(a,b,c,d)$ compared to what was described in Sec.~6.4 of Ref.~\cite{Bally21b}.
Using the notations of the latter article, the matrix elements stored are such that
\begin{equation}
\label{eq:indiset}
\begin{split}
 a &\in \llbracket 1, d_{\mathcal{M}} \rrbracket , \\
 b &\in \llbracket a+1, d_{\mathcal{M}} \rrbracket , \\
 c &\in \llbracket a, d_{\mathcal{M}} \rrbracket , \\
 d &\in \llbracket c+1, d_{\text{max}} \rrbracket \text{ with } \left\{ 
 \begin{array}{l}
   d_{\text{max}} = b \text{ if } c = a ,  \\
   d_{\text{max}} = d_{\mathcal{M}} \text{ otherwise.}
 \end{array} \right.
\end{split}
\end{equation}
Note that this change also applies to the recent versions of \TAURUSvap.

\subsection{Dependencies}
\label{sec:dep1}

The code requires the \textsf{BLAS} and \textsf{LAPACK} libraries.
In addition, it is recommanded to use a recent compiler as the code includes a few Fortran 2003/2008 commands that might not be implemented in compilers that are too old.

\section{Using the code \TAURUSmix}
\label{sec:structmix}

As a preliminary warning, we would like to stress the fact that the development of a software is an evolving process and, therefore, part of the information given here may become obsolete in future versions of the code.

\subsection{Source files and compilation}
The numerical code \TAURUSmix~is divided into different files containing the main program, the various modules, and specific mathematical routines.
The list all the files, in the order of which they should be compiled, is the following: \\
$\tt module\_constants.f90       $ \\
$\tt module\_mathmethods.f90     $ \\ 
$\tt module\_parameters.f90      $ \\
$\tt module\_cutoffs.f90         $ \\
$\tt module\_projmatelem.f90     $ \\
$\tt module\_spectroscopy.f90    $ \\
$\tt module\_initialization.f90  $ \\
$\tt taurus\_mix.f90             $ 

On the GitHub repository, we provide a makefile to compile the code that takes in entry one argument: \texttt{FC} to specify the compiler. If no argument is entered, the script will use some default values.
The correct values for the argument \texttt{FC} are: \texttt{gfortran} (default) and \texttt{ifort}. 
For example, if one wants to compile \TAURUSmix~using the \texttt{gfortran} compiler, one has to execute the command
\begin{equation*}
\texttt{make FC=gfortran}
\end{equation*}

For completeness, we also provide in the repository a bash script that offers comparable compilation capabilities. 
More information about how to use this script or the makefile can be found in the file \texttt{README.md} present in the main directory
of the repository.

\subsection{Execution}
Once in the directory containing the executable file, the code can be run by typing the command
\begin{equation*}
\texttt{./taurus\_mix.exe < input.txt}
\end{equation*}
where \texttt{input.txt} is the STDIN file containing the input parameters. The details concerning the format of STDIN can be found in the 
file \texttt{manual\_input.pdf} present in the directory \texttt{doc}. The code also requires, in the same directory, the files containing the matrix elements of the various operators as produced in output of a successful run of \TAURUSpav. See the file \texttt{doc/manual\_input.pdf} for more details.

To simplify the execution of the code, we provide the script \texttt{launch.sh} that performs all the necessary steps to run a calculation. 
To use it, go to the main directory of your copy of the repository and type the command 
\begin{equation*}
\texttt{bash launch.sh}
\end{equation*}

During its execution, the code prints various information (projected states read and kept after applying the cutoffs, list of the the norm eigenvalues, etc.) in the STDOUT. 
We recommend to store the printing in a file, for example \texttt{output.txt}, by typing 
\begin{equation*}
\texttt{./taurus\_mix.exe < input.txt > output.txt}
\end{equation*}
or
\begin{equation*}
\texttt{bash launch.sh > ouput.txt}
\end{equation*}

Additionally, the code will produce other files containing relevant information such as the collective wave functions or the energy spectrum ordered by ascending excitation energy or spin-parity.
The names of all the files produced during a run are recalled in the STDOUT.

\subsection{Dependencies}
\label{sec:dep2}

The code requires the \textsf{BLAS} and \textsf{LAPACK} libraries.
In addition, it is recommanded to use a recent compiler as the code includes a few Fortran 2003/2008 commands that might not be implemented in compilers that are too old.

%
%
\begin{acknowledgements}
We thank H.~Hergert for providing us with the matrix elements of the interaction as well as A.~Scalesi for generating the normal-ordered interactions for $^{24}$Mg. In addition, we thank A.~S{\'a}nchez-Fern{\'a}ndez, J.M.~Yao and M.~Frosini for helping us test or benchmark preliminary versions of the codes. This work was performed using HPC resources from GENCI-TGCC (Contract No.\ A0150513012) and CCRT (TOPAZE supercomputer) at Bruyères-le-Châtel.
This project has received funding from the European Union’s Horizon 2020 research and innovation programme under 
the Marie Skłodowska-Curie grant agreement No.~839847.
\end{acknowledgements}
%
%
\appendix
%
%
\bibliographystyle{apsrev}

\bibliography{biblio}
%
%
\end{document}